\documentclass[12pt]{article}
\usepackage{latexsym}
\usepackage{epsfig}
\usepackage{amsmath}

\begin{document}
\vspace*{-.6in}
\thispagestyle{empty}
\begin{flushright}
CALT-68-2525
\end{flushright}
\baselineskip = 18pt

\vspace{1.5in} {\Large
\begin{center}
Superconformal Chern--Simons Theories\end{center}} \vspace{.5in}

\begin{center}
John H. Schwarz
\\
\emph{California Institute of Technology\\ Pasadena, CA  91125,
USA}
\end{center}
\vspace{1in}

\begin{center}
\textbf{Abstract}
\end{center}
\begin{quotation}
\noindent We explore the possibilities for constructing Lagrangian
descriptions of three-dimensional superconformal classical gauge
theories that contain a Chern--Simons term, but no kinetic term,
for the gauge fields. Classes of such theories with ${\cal N} =1$
and ${\cal N} =2$ supersymmetry are found. However, interacting
theories of this type with ${\cal N} =8$ supersymmetry do not
exist.
\end{quotation}

\newpage

\pagenumbering{arabic}

\section{Introduction}

Many examples of conformal field theories are known in two
dimensions and in four dimensions. However, much less is known in
three dimensions. From the perspective of AdS/CFT one is
particularly interested in conformally invariant gauge theories,
where the rank of the gauge group is related to the amount of flux
in the dual AdS description. M theory admits compactifications
involving $AdS_4$, the most symmetrical choice being $AdS_4 \times
S^7$ \cite{Freund:1980xh}. According to the AdS/CFT conjecture
\cite{Maldacena:1997re} this should be dual to a three-dimensional
gauge theory with the superconformal symmetry $OSp(8|4)$. This
gauge theory should have gauge group $U(N)$ if the dual M theory
background has $N$ units of flux through the seven-sphere. The
situation ought to be rather analogous to the case of type IIB
superstring theory compactified on $AdS_5 \times S^5$, with $N$
units of flux, for which the dual gauge theory is ${\cal N} = 4$
supersymmetric Yang--Mills theory with a $U(N)$ gauge group and
the superconformal symmetry is $PSU(4|2,2)$. There are some
significant differences, however. For one thing the type IIB
superstring background contains a constant dilaton field, whose
value corresponds to the Yang--Mills coupling constant. There is
no analogous scalar field in the M theory case. Therefore the dual
three-dimensional CFT should not have an adjustable coupling, and
therefore it is expected to be strongly coupled. This makes it a
logical possibility that there is no explicit Lagrangian
description of this theory, but it does not imply that this must
be the case.

The usual viewpoint, which surely is correct, is the following:
The low-energy effective world-volume theory on a collection of
$N$ coincident D2-branes of Type IIA superstring theory is a
maximally supersymmetric $U(N)$ Yang-Mills theory in three
dimensions. This theory, which is not conformal because the
Yang-Mills coupling in three dimensions is dimensionful, has an
$SO(7)$ R symmetry corresponding to rotations of the transverse
directions. In the flow to the infrared the gauge coupling
increases, which corresponds to the string coupling (the vev of
the dilaton) increasing. This in turn corresponds to the radius of
the circular 11th dimension increasing. In the limit that the
coupling becomes infinite, one reaches the conformally-invariant
fixed point theory that describes a collection of coincident
M2-branes in eleven dimensions. This theory should have an
enhanced $SO(8)$ R symmetry corresponding to rotations of the
eight transverse dimensions. One question that we wish to explore
in this paper is whether it is possible to find an alternative
characterization of this fixed-point theory with an explicit
classical Lagrangian.

One can anticipate the field content of these theories from the
relation to M2-branes (in the M theory case) and D3-branes (in the
type IIB superstring theory case). The world-volume field content
of a single D3-brane contains a vector, six scalars, and four
Majorana spinors. To describe $N$ coincident D3-branes (at low
energy) it is just a matter of promoting these to $N\times N$
Hermitian matrices and constructing an interacting superconformal
field theory with $U(N)$ gauge symmetry. This is achieved by
${\cal N} =4$ SYM theory, of course. In the case of an M2-brane,
the physical world-volume field content consists of eight scalars
and eight (two-component) Majorana spinors. So a natural guess is
that these should be made into $N \times N$ matrices and the
$U(N)$ global symmetry should be gauged. However, this is not
entirely obvious, because unlike the case of D-branes, there is no
simple interpretation in terms of strings stretched between the
branes. When viewed in terms of the maximally supersymmetric SYM
theory that flows to the desired fixed-point theory one sees this
field content except that one of the matrix scalars is replaced by
a propagating gauge field. In the abelian case these can be
related by a duality transformation, but in the nonabelian case
there is no simple way of doing that. Rather than trying to carry
out such a duality transformation, we will start with the
postulated field content, which is clearly required for exhibiting
the desired $Spin(8)$ R symmetry.

The $U(N)$ gauge theory should have ${\cal N} = 8$
super-Poincar\'e symmetry and scale invariance, which together
ought to imply the full $OSp(8|4)$ superconformal
symmetry.\footnote{Usually, but not always, Poincar\'e invariance
together with scale invariance implies conformal invariance. (See
\cite{Awad:2000ie} and references therein.)} If one succeeds in
constructing such a theory, then it would be reasonable to expect
that quantum corrections do not destroy the scale invariance, like
in the case of ${\cal N}=4$ SYM theory.

The scalars and spinors in the proposed three-dimensional CFT give
an equal number of physical bosonic and fermionic degrees of
freedom. Therefore, to maintain supersymmetry when $U(N)$ gauge
fields are added, the number of bosonic degrees of freedom should
not change. This should be contrasted with the case of ${\cal N}
=4$ SYM theory, where the transverse polarizations of the gauge
fields are required to achieve an equal number of bosonic and
fermionic degrees of freedom. Starting from the free theory with
global $U(N)$ symmetry in three dimensions there are three
alternative ways to introduce the gauge fields that one might
consider: (1) Add gauge field couplings to make the global $U(N)$
symmetry local, but do not introduce kinetic terms for the gauge
fields. (2) Add gauge field couplings to make the global $U(N)$
symmetry local and add $F^2$ kinetic terms for the gauge fields.
(3) Add gauge field couplings to make the global $U(N)$ symmetry
local and add a Chern--Simons term for the gauge fields. We claim
that choice number (1) is inconsistent with supersymmetry, because
the gauge fields would give rise to constraints that would
effectively subtract bosonic degrees of freedom. Similarly, choice
number (2) is unacceptable, because the gauge fields would add
bosonic degrees of freedom. Also $F^2$ is dimension 4, and scale
invariance of the classical theory only allows dimension 3 terms.
This leaves choice (3), which I claim is exactly right. The
Chern--Simons term is dimension 3 and its inclusion does not lead
to either an increase or a decrease in the number of propagating
bosonic degrees of freedom, so it is conceivable that
supersymmetry can be achieved.

To be honest, it is quite mysterious how a Chern--Simons term
could be generated in the IR flow of the SYM theory discussed
earlier. This is especially a concern since the SYM theory that
flows to the fixed point in question is parity conserving. So how
could the theory be parity violating at the fixed point. In the
end, we will not find such an ${\cal N} =8$ theory, and maybe this
is one of the reasons why.

As we have said, the problem that we would most like to solve is
the explicit construction of a Lagrangian for the
three-dimensional CFT that has maximal supersymmetry and is dual
to M theory on $AdS_4 \times S^7$. However, most of this paper
will address more modest goals: the construction of
three-dimensional gauge theories with ${\cal N} = 1$ and ${\cal N}
= 2$ supersymmetry and classical scale invariance. This will
provide a framework for explaining why an ${\cal N} = 8$ super
Chern--Simons theory cannot be constructed. However, it is
conceivable that one could construct a Lagrangian description of
the desired $OSp(8|4)$ superconformal theory by modifying one or
more of our assumptions.

\section{Supersymmetry of Chern-Simons Theories}

Pure Chern--Simons theory has a Lagrangian that is proportional to
\begin{equation}
{\cal L}_{\rm CS} = {\rm tr} \Big[ \epsilon^{\mu\nu\rho}(A_{\mu}
\partial_{\nu} A_{\rho} + \frac{2i}{3} A_{\mu}A_{\nu} A_{\rho}) \Big] .
\end{equation}
It gives the classical field equation
\begin{equation}
F_{\mu\nu} = \partial_{\mu} A_{\nu} - \partial_{\nu} A_{\mu} +i
[A_{\mu}, A_{\nu}] =0.
\end{equation}
A curious fact about this theory is that it has any desired amount
of supersymmetry, if one simply decrees that $A_{\mu}$ is
invariant under each of the supersymmetry transformations. The
reason this is possible is that this theory has no propagating
on-shell degrees of freedom. To prove this assertion one needs to
verify the super-Poincar\'e algebra, especially that the
commutator of two supersymmetry transformations is a translation.
Since the supersymmetry transformation is trivial, this means that
the translation symmetry transformation should also be trivial.

Since $A_{\mu}$ by itself is certainly not a complete off-shell
supermultiplet, the supersymmetry algebra should only hold
on-shell. This means that in verifying the closure of the algebra,
one is allowed to use the field equation $F_{\mu\nu}=0$. This is a
familiar situation; many of the nicest supersymmetric theories,
such as ${\cal N} = 4$ super Yang--Mills theory, do not have a
straightforward formulation in terms of off-shell supermultiplets
of the full supersymmetry algebra. So the proof of the assertion
that we are making is simply to show that an infinitesimal
translation by a constant amount $a^{\rho}$ is trivial modulo a
gauge transformation and the equations of motion. This is the case
because an infinitesimal translation shifts $A_{\mu}$ by
$a^{\rho}\partial_{\rho} A_{\mu}$, which differs from $a^{\rho}
F_{\rho \mu}$ by an infinitesimal gauge transformation
\begin{equation}
\delta A_{\mu} = \nabla_{\mu}\Lambda = \partial_{\mu} \Lambda + i
[A_{\mu}, \Lambda]
\end{equation}
for the choice $\Lambda = a^{\rho} A_{\rho}$. This then vanishes
by the equations of motion. Of course, this triviality of
translation invariance is not a big surprise since Chern--Simons
is a topological theory.

We will be interested in coupling the Chern--Simons gauge field to
other fields. For this purpose it is convenient to have complete
off-shell supermultiplets. This enables one to combine
supersymmetric expressions without substantial modification of the
supersymmetry transformation formulas, as we will see. Pure
Chern--Simons theories with off-shell supersymmetry were
constructed in~\cite{Nishino:1991sr} for ${\cal N} = 1,2,4$. That
work did not discuss coupling these supermultiplets to other
matter supermultiplets.

\section{${\cal N} = 1$ Models}

\subsection{The Gauge Multiplet}

One of the nice things about the ${\cal N} = 1$ theories in three
dimensions that we want to construct is that it is easy to
implement supersymmetry by using superfields. The Grassmann
coordinates of  ${\cal N} = 1$ superspace consist just of a
two-component Majorana spinor. There are two kinds of multiplets
that we will be interested in: gauge multiplets and scalar
multiplets. In this section we discuss the gauge multiplet. This
superfield is a spinor. However, in this case we find it
convenient to work with the component fields that survive in the
three-dimensional analog of Wess--Zumino gauge. These are the
gauge field $A_{\mu}$ and a Majorana two-component spinor $\chi$.
Both of these are in the adjoint representation of the Lie algebra
and can be represented as Hermitian matrices in some convenient
representation, which will be specified later when they are
coupled to scalar supermultiplets.

Since we are mainly interested in classical considerations in this
paper, we will not specify the overall normalization of the action
at this time. This would need to be considered carefully in
defining the quantum theory, of course. With this understanding we
choose the ${\cal N} = 1$ Chern--Simons Lagrangian to be
\begin{equation} \label{CS}
{\cal L}_{\rm CS} = {\rm tr} \Big[ \epsilon^{\mu\nu\rho}(A_{\mu}
\partial_{\nu} A_{\rho} + \frac{2i}{3} A_{\mu}A_{\nu} A_{\rho}) -
\bar\chi \chi \Big] .
\end{equation}
This theory differs from the pure Chern--Simons theory discussed
in the preceding section only by the addition of the auxiliary
fermi field $\chi$. Note that the Lagrangian has dimension three
for the choices dim $A =1$ and dim $\chi = 3/2$, and then the
action is scale invariant.

To get off-shell closure of the supersymmetry algebra, one needs
to have an equal number of off-shell bosonic and fermionic degrees
of freedom. In fact, taking account of gauge invariance, $A_{\mu}$
and $\chi$ both have two off-shell modes. This ensures off-shell
closure of the supersymmetry algebra without use of equations of
motion.

The infinitesimal supersymmetry transformations that leave ${\cal
L}_{\rm CS}$ invariant (up to a total derivative) are\footnote{For
the most part, we follow the conventions of
ref.~\cite{Marcus:1983hb}. The metric has signature $+--$ and
$\bar\chi = \chi^T \gamma^0$. (The transpose here only acts on the
spinor components and not on the Lie algebra matrix.) A possible
choice of the Dirac matrices in terms of standard Pauli matrices
is $\gamma^0 =\sigma_2$, $\gamma^1 = i\sigma_3$, and $\gamma^2 =
i\sigma_1$. Note that then $\gamma^{\mu\nu\rho} =-i
\epsilon^{\mu\nu\rho}$ and $\gamma^{\mu\nu} = -i
\epsilon^{\mu\nu\rho} \gamma_{\rho}$.}
\begin{equation}
\delta A_{\mu} = i \bar\varepsilon \gamma_{\mu} \chi
\end{equation}
and
\begin{equation}
\delta \chi = \frac{1}{2} \gamma^{\mu\nu} F_{\mu\nu}\,
\varepsilon.
\end{equation}
The commutator of two supersymmetry transformations $[\delta_1,
\delta_2]$ gives the sum of a spacetime translation by $a^{\rho} =
2i \bar\varepsilon_1 \gamma^{\rho} \varepsilon_2$ and a gauge
transformation by $\Lambda = - a^{\rho} A_{\rho}$.

\subsection{The Matter Theory}

Let us now turn to the scalar supermultiplets, which we write in
terms of superfields as follows:
\begin{equation}
\Phi = \phi + \bar\theta\psi + \frac{1}{2} \bar\theta\theta C.
\end{equation}
Let us take $\Phi^a$, $a=1,2,\ldots, {\rm dim}\, R$, to belong to
a representation $R$ of the gauge group $G$. We may assume without
loss of generality that $R$ is real. Then there is no need to make
a distinction between upper and lower indices. In this section we
will formulate a scale invariant theory of the scalar superfields
with global $G$ symmetry. In the next section we will couple this
theory to the gauge supermultiplet, so as to achieve local $G$
symmetry while retaining global ${\cal N} = 1$ supersymmetry.

To achieve scale invariance we assign dimension 1/2 to $\Phi$.
This implies that dim $\phi =1/2$, dim $\psi =1$, and dim $C
=3/2$. Then the most general scale-invariant theory is given by
the $\bar\theta \theta$ component of a dimension two superfield
expression. The only possibilities are a kinetic term of the form
$\bar D \Phi^a D \Phi^a$, where $D$ is the usual supercovariant
derivative, and an interaction term of the form $W = t_{abcd}
\Phi^a \Phi^b \Phi^c \Phi^d$. The dimensionless symmetric tensor
$t_{abcd}$ is restricted by the requirement of $G$ invariance. In
terms of component fields we obtain the matter Lagrangian
\begin{equation} \label{Lm0}
{\cal L}_{\rm m}^0 = \frac{1}{2} \partial_{\mu} \phi^a
\partial^{\mu} \phi^a +\frac{i}{2} \bar\psi^a
\gamma^{\mu}\partial_{\mu} \psi^a + \frac{1}{2} C^a C^a + t_{abcd}
\phi^a \phi^b \big(\frac{1}{3}\phi^c C^d - \frac{1}{2}\bar\psi^c
\psi^d \big).
\end{equation}
Note that elimination of the auxiliary field $C$ would give a term
of the structure $\phi^6$.

The supersymmetry transformations that leave this Lagrangian
invariant (up to a total derivative) are
\begin{equation}
\delta \phi^a = \bar\varepsilon \psi^a = \bar\psi^a \varepsilon
\end{equation}
\begin{equation}
\delta\psi^a = -i \gamma^{\mu} \varepsilon \partial_{\mu} \phi^a +
C^a \varepsilon
\end{equation}
\begin{equation}
\delta C^a = -i \bar\varepsilon \gamma^{\mu}\partial_{\mu} \psi^a
=i \partial_{\mu} \bar\psi^a\gamma^{\mu} \varepsilon.
\end{equation}
This algebra also has off-shell closure giving a translation with
the same parameter as in the case of the gauge supermultiplet.

When this system is coupled to the gauge supermultiplet, so as to
achieve local $G$ symmetry and global supersymmetry, the
supersymmetry transformations of the gauge supermultiplet are
unchanged and the supersymmetry transformations of the matter
multiplets get a few additional terms (described in the next
section) that are required for them to be covariant. Then the
commutator of two supersymmetry transformations of the matter
fields gives rise to a gauge transformation as well as a
translation, with the same parameters as in the case of the gauge
supermultiplet discussed earlier.

\subsection{The Gauged Theory}

We can now put the ingredients together to define the most general
gauge-invariant ${\cal N} =1$ theory that has classical scale
invariance. For this purpose it is convenient to represent the
gauge fields by matrices $(A_{\mu})^a{}_b$ and $\chi^a{}_b$ in the
representation $R$ of the Lie algebra. The total Lagrangian is
${\cal L} = {\cal L}_{\rm CS} + {\cal L}_{\rm m}$, where ${\cal
L}_{\rm CS}$ is given in eq.~(\ref{CS}) and ${\cal L}_{\rm m}$ is
eq.~(\ref{Lm0}) embellished by couplings to the gauge
supermultiplet. The gauged matter Lagrangian takes the form
\begin{equation} \label{Lm}
{\cal L}_{\rm m} = \frac{1}{2} (\nabla_{\mu} \phi)^a (\nabla^{\mu}
\phi)^a +\frac{i}{2} \bar\psi^a \gamma^{\mu}(\nabla_{\mu} \psi)^a
+ \frac{1}{2} C^a C^a + i\phi^a \bar\chi^{ab} \psi^b + t_{abcd}
\phi^a \phi^b [\frac{1}{3}\phi^c C^d - \frac{1}{2}\bar\psi^c
\psi^d]  ,
\end{equation}
where $\nabla_{\mu} \Phi^a =
\partial_{\mu} \Phi^a +i (A_{\mu})^{ab} \Phi^b$.

The supersymmetry transformations of the combined system are
\begin{equation}
\delta A_{\mu} = i \bar\varepsilon \gamma_{\mu} \chi
\end{equation}
\begin{equation}
\delta \chi = \frac{1}{2} \gamma^{\mu\nu} F_{\mu\nu}\,
\varepsilon.
\end{equation}
\begin{equation}
\delta \phi^a = \bar\varepsilon \psi^a
\end{equation}
\begin{equation}
\delta\psi^a = -i \gamma^{\mu} \varepsilon (\nabla_{\mu} \phi)^a +
C^a \varepsilon
\end{equation}
\begin{equation}
\delta C^a = -i \bar\varepsilon \gamma^{\mu} (\nabla_{\mu} \psi)^a
+ i \bar\varepsilon \chi^{ab} \phi^b.
\end{equation}
The only change from before is the replacement of ordinary
derivatives by covariant derivatives and the addition of the
second term in $\delta C^a$.

\section{${\cal N} = 2$ Models}

Let us now try to find models that have ${\cal N} = 2$
supersymmetry. Since ${\cal N} = 2$ in three dimensions is closely
related to ${\cal N} = 1$ in four dimensions, and also has a
$U(1)$ R symmetry, a complex notation is convenient. For previous
related work see \cite{Gates:1983nr} \cite{Zupnik:1988en}
\cite{Ivanov:1991fn}.

\subsection{The Gauge Multiplet}

The Chern--Simons part of the action is constructed out of a
vector supermultiplet that can be obtained by dimensional
reduction of a four-dimensional ${\cal N} =1 $ supermultiplet. In
four dimensions the multiplet contains a gauge field $A_{\mu}$, a
four-component Majorana spinor $\chi$, and a real scalar $D$. On
reduction to three dimensions, the gauge field gives a
three-vector gauge field $A_{\mu}$ and a scalar $\sigma$,
corresponding to the component $A_3$ in four dimensions. The
spinor can be recast as a two-component Dirac spinor $\chi$, and
we still have the scalar $D$. Off-shell there are four bosonic and
four fermionic degrees of freedom. In the Chern--Simons theory
that we will construct, there are no propagating on-shell degrees
of freedom.

Note that the dimension of $A_{\mu}$ and $\sigma$ is 1, the
dimension of $\chi$ is $3/2$, and the dimension of $D$ is 2. In
terms of an infinitesimal Dirac spinor $\varepsilon$, the
supersymmetry transformations for a nonabelian gauge multiplet are
the following
\begin{equation}
\delta A_{\mu} = \frac{i}{2}( \bar\varepsilon \gamma_{\mu} \chi -
\bar\chi \gamma_{\mu} \varepsilon )
\end{equation}
\begin{equation}
\delta \sigma = \frac{i}{2}( \bar\varepsilon  \chi - \bar\chi
\varepsilon )
\end{equation}
\begin{equation}
\delta D = \frac{1}{2}( \bar\varepsilon \gamma^{\mu}
\nabla_{\mu}\chi + \nabla_{\mu}\bar\chi \gamma^{\mu} \varepsilon )
+ \frac{i}{2}( \bar\varepsilon  [\chi, \sigma] + [\bar\chi,
\sigma] \varepsilon )
\end{equation}
\begin{equation}
\delta \chi = \big(\frac{1}{2} \gamma^{\mu\nu} F_{\mu\nu} - iD
-\gamma^{\mu} \nabla_{\mu} \sigma\big) \varepsilon.
\end{equation}
The hermitian conjugate of the last formula is
\begin{equation}
\delta \bar\chi = \bar\varepsilon\big(-\frac{1}{2} \gamma^{\mu\nu}
F_{\mu\nu} +iD -\gamma^{\mu} \nabla_{\mu} \sigma\big) .
\end{equation}
The commutator of two supersymmetry transformations gives a
translation by an amount
\begin{equation}
a^{\rho} = i( \bar\varepsilon_1 \gamma^{\rho}\varepsilon_2 -
\bar\varepsilon_2 \gamma^{\rho}\varepsilon_1)
\end{equation}
and a gauge transformation with parameter
\begin{equation}
\Lambda = - a^{\rho} A_{\rho} + i(\bar\varepsilon_1 \varepsilon_2
- \bar\varepsilon_2 \varepsilon_1)\sigma .
\end{equation}

We can now construct a supersymmetric Chern--Simons action out of
this supermultiplet. The result is
\begin{equation} \label{CS2}
{\cal L}_{\rm CS} = {\rm tr} \Big[ \epsilon^{\mu\nu\rho}(A_{\mu}
\partial_{\nu} A_{\rho} + \frac{2i}{3} A_{\mu}A_{\nu} A_{\rho}) -
\bar\chi \chi +2D\sigma\Big] .
\end{equation}
Note that each of the terms is dimension three.

\subsection{The Matter Theory}

The notation now is that the index for the matter representation
$R$ of the gauge group $G$ is not displayed explicitly, but
another index $A$ labelling repetitions of $R$ is displayed. If
the matter representation $R$ is complex, let us use the notation
$(\Phi^A)^{\star} = \Phi_A$ to distinguish holomorphic fields and
their antiholomorphic conjugates. These can be identified as
three-dimensional counterparts of chiral and antichiral
superfields in four dimensions. The multiplet contains a complex
scalar $\phi^A$ of dimension $1/2$, a Dirac two-component spinor
$\psi^A$ of dimension 1, and a complex auxiliary scalar $F$ of
dimension $3/2$. We also have the following R charge assignments:
$\phi^A$ has R charge 1/2, $\psi^A$ has R charge $-1/2$, and $F^A$
has R charge $-3/2$. (The conjugates take the negatives of these
values, of course.) These conventions correspond to the
holomorphic superspace coordinate $\theta$ having R charge 1, and
the supersymmetry parameter $\varepsilon$ having R charge $-1$.

We can now write down the supersymmetry transformations for this
multiplet. They are
\begin{equation}
\delta \phi^A = \bar\varepsilon \psi^A
\end{equation}
\begin{equation}
\delta\psi^A = -i \gamma^{\mu}  \partial_{\mu} \phi^A \varepsilon
+ F^A \varepsilon^{\star}
\end{equation}
\begin{equation}
\delta F^A = -i \bar\varepsilon^{\star} \gamma^{\mu}\partial_{\mu}
\psi^A .
\end{equation}
These formulas are determined, up to coefficients, by dimensional
analysis and R symmetry.  One can verify that the supersymmetry
algebra closes off-shell giving the same translation parameter as
for the gauge multiplet.

The ${\cal N} =2$ supersymmetric matter Lagrangian takes the form
\begin{equation} \label{Lm2}
{\cal L}_{\rm m} =  \partial_{\mu} \phi_A \partial^{\mu} \phi^A +i
\bar\psi_A \gamma^{\mu} \partial_{\mu} \psi^A + F_A F^A + W_F +
W_F^{\star}.
\end{equation}
Here $W_F$ and $ W_F^{\star}$ represent superpotential F-terms,
which need to be quartic for scale invariance. As before, they
give terms of the form $\phi^2\psi^2$ and $\phi^3 F$. The overall
normalization of the Lagrangian is arbitrary.

\subsection{The Gauged Theory}

In the gauged theory the supersymmetry transformations of the
matter supermultiplet take the form
\begin{equation}
\delta \phi^A = \bar\varepsilon \psi^A
\end{equation}
\begin{equation}
\delta\psi^A = (-i \gamma^{\mu}  \nabla_{\mu} \phi^A -\sigma
\phi^A )\varepsilon + F^A \varepsilon^{\star}
\end{equation}
\begin{equation}
\delta F^A = \bar\varepsilon^{\star} (-i\gamma^{\mu}\nabla_{\mu}
\psi^A +i\chi \phi^A + \sigma \psi^A).
\end{equation}
For these rules the commutator of two supersymmetry
transformations gives a translation and a gauge transformation
with the same parameters as for the gauge supermultiplet.

The matter Lagrangian that is invariant (up to a total derivative)
under these transformations is
\begin{equation} \label{Lm2g}
{\cal L}_{\rm m} =  (\nabla_{\mu} \phi)_A (\nabla^{\mu} \phi)^A +i
\bar\psi_A \gamma^{\mu} (\nabla_{\mu} \psi)^A + F_A F^A
\end{equation}
\[
-\phi_A \sigma^2 \phi^A +\phi_A D \phi^A -\bar\psi_A \sigma \psi^A
+i \phi_A \bar\chi \psi^A -i\bar\psi_A \chi\phi^A  + W_F +
W_F^{\star}.
\]
Combining this with the Chern--Simons terms in eq.~(\ref{CS2}), it
is straightforward to eliminate the auxiliary fields $\sigma$,
$D$, $\chi$, and $F$. This gives rise to various $\phi^2\psi^2$
and $\phi^6$ terms.

\section{The ${\cal N} = 8$ Theory?}

The free $U(1)$ theory that is the low-energy effective
world-sheet theory of an M2-brane in 11 dimensions is well-known
\cite{Bergshoeff:1987cm} \cite{Bergshoeff:1987qx}. The matter
field content consists of scalars $\phi^I$ in the $8_{\rm v}$
representation of $Spin(8)$ and Majorana spinors $\psi^A$ in the
$8_{\rm s}$ representation. The eight supersymmetries belong to
the $8_{\rm c}$ representation, and the parameters can be denoted
$\varepsilon^{\dot A}$. The assignment of these representations is
arbitrary, because of triality symmetry. However, the association
of the scalars with the vector representation is a natural choice,
because they describe excitations of the M2-brane in the eight
transverse directions. If one were to add a decoupled $U(1)$ gauge
field described by a Chern--Simons action, this would be rather
inconsequential, since it has no propagating degrees of freedom
and is supersymmetric by itself, as was explained in section 2.

The free matter Lagrangian is
\begin{equation} \label{Lm8}
{\cal L}_{\rm m} =  \partial_{\mu} \phi^I
\partial^{\mu} \phi^I +i \bar\psi^A \gamma^{\mu} \partial_{\mu}
\psi^A .
\end{equation}
This is invariant (up to total derivatives) under the
supersymmetry transformations
\begin{equation}
\delta \phi^I = \bar\varepsilon^{\dot A} \Gamma^I_{\dot A A}\psi^A
\end{equation}
\begin{equation}
\delta\psi^A = -i \Gamma^I_{A \dot A}\gamma^{\mu}  \partial_{\mu}
\phi^I \varepsilon^{\dot A},
\end{equation}
where $\Gamma^I_{\dot A A}$ and its transpose are invariant
tensors that describe the coupling of the three 8s of $Spin(8)$.
This is the same structure as in the two-dimensional light-cone
gauge world-sheet action for the type IIB superstring in the GS
formalism. (The IIA theory uses different representations for
left-movers and right-movers.)

The problem now is to find the three-dimensional theory that
describes $N$ coincident M2-branes and is the CFT dual of M theory
on $AdS_4 \times S^7$ with $N$ units of flux through the
seven-sphere. By analogy with the duality between ${\cal N} =4$
SYM theory and type IIB superstring theory on $AdS_5 \times S^5$,
it is natural to expect that we need a $U(N)$ gauge theory (in
which the $U(1)$ component decouples) with the matter fields in
the adjoint representation. This would be consistent with the idea
that they are part of the same supermultiplet as the gauge fields.
In that case, it is convenient to represent them by $N \times N$
hermitian matrices. There are reasons for concern, however. One is
that the number of degrees of freedom should scale as $N^{3/2}$
for large $N$ \cite{Klebanov:1996un}, whereas the type of
construction we are contemplating would appear to give an $N^2$
scaling. Another concern is the parity-conservation issue
described in the introduction. We proceed nonetheless with the
justification that the existence or nonexistence of an ${\cal N} =
8$ $U(N)$  Chern--Simons theory with the indicated field content
is of intrinsic interest irrespective of any possible
applications.

A possible approach for constructing the interacting ${\cal N} =
8$ theory is to specialize the ${\cal N} = 2$ results obtained
above to the case where the representation $R$ consists of four
complex copies of the adjoint representation. This gives the right
field content, and it allows us to make an $SU(4)$ global symmetry
(in addition to the $U(1)$ R symmetry) manifest. Once this is
achieved, we can try to establish the full ${\cal N} = 8$
structure with its $Spin(8)$ R symmetry. This approach is
analogous to formulating ${\cal N} = 4$ SYM theory in terms of
${\cal N} = 1$ superfields. In that formulation only a $U(1)\times
SU(3)$ subgroup of the full $SU(4)$ R symmetry is manifest.

In the ${\cal N} = 4$ SYM construction there is a superpotential
$W = \lambda\, \epsilon_{ABC} {\rm tr} (\Phi^A \Phi^B \Phi^C)$.
When the coefficient $\lambda$ is given the appropriate value, the
manifest $SU(3) \times U(1)$ symmetry extends to $SU(4)$, and one
obtains ${\cal N} =4$ SYM. In the present problem it seems
reasonable to expect an analogous story in which the manifest
$SU(4) \times U(1)$ symmetry extends to $Spin(8)$. The
superpotential is constructed out of four superfields in the 4 of
$SU(4)$. So the analogous superpotential would seem to be
$\lambda\, \epsilon_{ABCD} {\rm tr} (\Phi^A \Phi^B \Phi^C
\Phi^D)$. Unfortunately, this vanishes due to the conflicting
symmetries of the trace and the epsilon symbol.

The only nonzero possibility appears to be $\epsilon_{ABCD} {\rm
tr} (\Phi^A) {\rm tr} (\Phi^B \Phi^C \Phi^D)$. Such a formula
would imply that the singlet component of the $U(N)$ fields
couples nontrivially. This conflicts with the structure of the
rest of the theory, as well as with all expectations. Thus it
appears that there cannot be a superpotential. However, without a
superpotential contribution, the rest of the theory does not have
the desired $SO(8)$ symmetry. This argument constitutes rather
strong evidence against the existence of an ${\cal N} = 8$ theory,
at least within the general framework that is being considered
here. However, as a check, the problem was also analyzed in terms
of component fields with the same conclusion.

\section{Discussion}

We have constructed a class of scale-invariant three-dimensional
gauge theories with ${\cal N} = 1$ and ${\cal N} = 2$
supersymmetry, which may be of some interest. For example, gauge
theories with three-dimensional conformal invariance could have
condensed matter applications \cite{Duval:1997pe}. However, our
main goal, the construction of scale-invariant gauge theories with
${\cal N} = 8$ supersymmetry, has not been achieved. There should
be a superconformal dual to M theory on $AdS_4\times S^7$, but
since the desired properties are only required at strong coupling,
realized as a nontrivial IR fixed point
\cite{Seiberg:1996nz}\cite{Intriligator:1996ex}\cite{Seiberg:1997ax},
there need not be  a classical lagrangian description.

If ${\cal N} = 8$ theories of the type that were sought had been
shown to exist, there are some interesting questions concerning
the AdS/CFT duality that would have arisen. One is the parity
issue discussed in the introduction: Chern--Simons theories are
parity violating, whereas the super Yang--Mills theory which is
supposed to flow to the desired conformal field theory in the IR
is not parity violating. Also, M theory is parity conserving.
Another concerns the level of the quantum Chern--Simons theory.
The ${\cal N} = 8$ gauge theory would be characterized by two
integers: $N$ (the rank of the gauge group) and $k$ (the
Chern--Simons level). The level $k$ is expected to be an integer,
because the boundary of the Euclideanized M theory geometry is a
three-sphere. The gauge coupling would be $g^2 \sim 1/k$. There
would be no other continuous parameters. However, two integers is
already more than is expected, because the dual geometry is
characterized entirely by one integer, the flux through the
seven-sphere.

An interesting possibility is that superconformal Chern--Simons
theories of the type described here could be dual to $AdS_4 \times
K$ compactifications of the massive (Romans) variant of type IIA
superstring theory \cite{Romans:1985tz}.\footnote{I am grateful to
Nikita Nekrasov and Greg Moore for this suggestion.}  The
quantized Romans' mass should correspond to the Chern--Simons
level. $AdS_4 \times K$ solutions of massive type IIA supergravity
with ${\cal N} = 1,2,4$ supersymmetry are described in
\cite{Behrndt:2004mj} \cite{Nunez:2001pt}. To pursue this one
would also want to study the superconformal symmetry of any
proposed dual Chern--Simons theories at the quantum level. (For
references on renormalization properties of Chern--Simons theories
see \cite{Avdeev:1992jt} \cite{Kapustin:1994mt}
\cite{DelCima:1997pb}.)

In conclusion, we have constructed large classes of ${\cal N} = 1$
and ${\cal N} = 2$ classical superconformal Chern--Simons
theories, which may be of some interest, but there is no classical
${\cal N} = 8$ Lagrangian of this type. Moreover, it seems
reasonable that there should be no classical lagrangian
description of the conformal field theory that is dual to M theory
on an $AdS_4\times S^7$ background.

\section*{Acknowledgments}

I am grateful to Lars Brink, Iosef Bena, Jon Bagger, Nikita
Nekrasov, Malcolm Perry, and Anton Kapustin for discussions. I
also wish to acknowledge the hospitality of the Aspen Center for
Physics and the Rutgers high energy physics group, where parts of
this work were done. This work was partially supported by the U.S.
Dept. of Energy under Grant No. DE-FG03-92-ER40701.

\end{document}